\documentclass[
  ,draft            
  ]
  {aipproc}

\layoutstyle{8x11double}

\begin{document}

\title{Planetary nebulae and the chemical evolution of the galactic bulge}


\author{R.D.D. Costa, A.V. Escudero, W.J. Maciel}{
  address={IAG/USP, S\~ao Paulo, Brazil}
}

\begin{abstract}
 Electron temperatures, densities, ionic and elemental
 abundances of helium, nitrogen, oxygen, argon, sulfur and neon were derived for a sample of bulge 
planetary nebulae, representative of its intermediate mass population.
 Using these results as constraints, a model for the chemical evolution of the galactic bulge
 was developed. The results indicate that the best fit is achieved using a double-infall model,
 where the first one is a fast collapse of primordial gas and the second is slower and enriched
 by material ejected by the bulge itself during the first episode.
\end{abstract}

\maketitle


\section{Introduction}

Planetary nebulae can constitute important tools to study the chemical evolution
of the galactic bulge. As representative objects of its intermediate mass population, they
can provide accurate determinations of the abundances of light elements such as helium and
nitrogen produced
by progenitor stars of different masses. Besides, abundances of heavier elements such as
oxygen, sulfur and neon can be also derived. These elements are produced in more massive 
stars and represent the abundances in the interstellar medium at the 
progenitor formation epoch.

In this work spectrophotometric results derived for a sample of 57 bulge PNe are reported.
Electron temperatures, densities, ionic and elemental abundances of helium, nitrogen, 
oxygen, argon, sulfur and neon were derived for the sample. Using these observational 
results as constraints, a model for the chemical evolution of the bulge was developed.

\section{Observations and data reduction}


Observations used in this work were carried out in two different telescopes: 1.60 m
LNA (Laborat\'orio Nacional de Astrof\'\i sica - Bras\'opolis, Brazil) and 1.52 m ESO
(European Southern Observatory - La Silla, Chile). All objects were observed using 
Cassegrain spectrographs. At LNA a 300 l/mm grid with 4.4 \AA/pix dispersion was used
and at ESO a 600 l/mm grid, with 2.4 \AA/pix dispersion. In both cases long
slit of 2" width were adopted. Each night at least three spectrophotometric standard
stars were observed in order to secure flux calibration. 
Observational details can 
be found in Escudero and Costa (2001) and Escudero et al. (2004).


Data reduction and analysis was performed using IRAF routines for long slit spectra.
Interstellar extinction was derived based on the H$\alpha$/H$\beta$ ratio, and was 
corrected using the extinction curve of Fitzpatrick (1999),
which proved to produce better results for bulge objects when compared with other curves.
Quality of the reddening correction was checked by comparing dereddened and theoretical
H$\gamma$/H$\beta$ ratios.

Electron temperatures were derived from the following line ratios: [OIII]$\lambda\lambda$
4363/5007 and [NII]$\lambda\lambda$ 5754/6584, which define respectively "high" and
"low" ionization regions. Electron density was derived from the sulfur ratio [SII]$\lambda\lambda$
6716/6731, corresponding therefore to an average density to the whole nebula.

Helium I abundances were derived taking into account the collisional excitation correction. 
Ionic abundances were derived using the fits by Alexander and Balick (1997), and checked
using a three-level atom model, with good agreement between both techniques. 
Elemental
bundances were estimated using ionization correction factors, as described in details in 
the papers above mentioned.
Figure 1 displays the distribution of the derived oxygen abundances, compared to other data
from the literature, without objects in common.

\begin{figure}
  \includegraphics[height=.27\textheight]{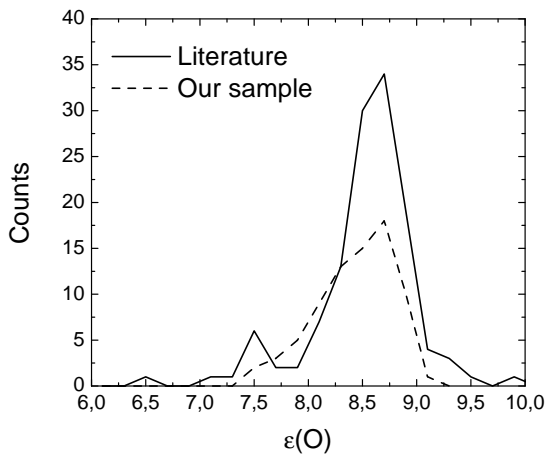}
  \caption{Distribution of oxygen abundances for bulge planetary nebulae}
\end{figure}

\section{Chemical evolution model}

A model was developed in order to describe the formation and chemical evolution of
the galactic bulge, using abundances derived from planetary nebulae as representative of its intermediate
age population.
 The results derived in the present work reinforce
some evolutive characteristics already derived in previous works, and present improvements
such as the inclusion of a second enriched infall to explain peculiarities of the abundance
distribution like the dispersion found in the correlation between nitrogen and oxygen
abundances.

\subsection{Basic ingredients}

The model was developed as a C++ code that simulates numerically the chemical evolution of the bulge,
using the basic ingredients to describe the evolution, namely stellar formation rate, initial mass
function, gas infall rate, stellar yields and mass distribution of SNIa progenitors:

\begin{itemize}
\item
Star formation rate: To determine the amount of interstellar gas converted to stars at each
time interval, the Schmidt law (Kennicutt 1998) was used. The adopted coefficient and exponent of the
law were the same derived by Kennicutt. For some runs, the stellar formation efficiency was
variated of two orders of magnitude to test its impact on the derived chemical abundances, without
significant differences.
 
\item
Initial mass function: basically, the initial mass function adopted was that of Kroupa (2002), which
reproduces satisfactorily the observed chemical abundances in planetary nebulae. However, the
O/Fe ratio for poor stars obtained from the literature (Pompeia et al. 2003) is higher than the values
derived from the model. To solve this problem, the Salpeter law was used for the first 0.3 Gyr
of the bulge formation, resulting in chemical abundances closer to those found in stars.

\item
Gas infall rate: the value of this rate along the Galaxy evolution is probably
the less known ingredient of a chemical evolution model. Since the published chemical evolution
models adopt a time-exponential infall rate, the same kind of expression was adopted in this
model. The only difference is the adopted gas infall scale, as well as its normalization constant,
that depends on the total mass of the bulge. This model consists in two gas infall episodes. For
the first one, in a time scale of 0.1 Gyr nearly all the bulge population was formed. For the 
second episode, a time scale of 2.0 Gyr was adopted, close to that derived for the inner disk
in chemical evolution models of the galactic disk.

\item
Stellar yields: the yields were divided in three categories: intermediate mass population,
SNII and SNIa. These yields were obtained respectively from van den Hoek \& Groenewegen (1997), 
Woosley \& Weaver (1995) and Tsujimoto et al. (1995). To check the results, the data from 
Tsujimoto et al. (1995) for SNII were also used, resulting in a higher O/Fe ratio, closer to those found
in the literature, however the abundance distribution does not change significantly with these
new yields. Figure 2 shows the derived distribution of oxygen abundances compared to model results
using two different stellar evolution models, and figure 3 shows the evolution of the O/Fe ratio 
using two yields for SNII.
 
\item
Mass distribution of the SNIa progenitors: SNIa have binary systems as progenitors. Their
mass distribution is not well known yet. A way to describe this distribution was 
provided by Chiappini (1997), and assumes that the highest probability for the mass ratio 
between primary and secondary stars in such a system is one. Since this distribution provides
satisfactory results for the disk abundances, it was adopted in the present model.
\end{itemize}

\subsection{Main characteristics of this model}

The model adopts a mixed scenario to describe the bulge evolution, with two main phases:
the first one is a fast collapse of primordial gas and the second consists in a slow collapse
of enriched material.

One of the improvements of the model presented here is the small amount of 
free parameters. Nearly all its ingredients can be checked observationally
and are already accepted in the literature. The adopted IMF was that of Kroupa (2002),
which is widely adopted in chemical evolution works. The adopted SFR is the Schmidt
law, with the constant and density exponent derived observationally by Kennicutt (1998).
The amount of binary systems generating SNIa was derived from solar neighbourhood data.
Finally, the stellar yields were derived from stellar evolution models (van den Hoek \&
Groenewegen 1997, Woosley \& Weaver 1995 and Tsujimoto et al. 1995). The only free
parameters are infall rate and wind rate. Unfortunately the gas infall and the gas loss/change
in the early evolution of the Galaxy are still not clearly described in the literature.
These are probably the two main points to be addressed in forthcoming chemical evolution models.

\begin{figure}
  \includegraphics[height=.27\textheight]{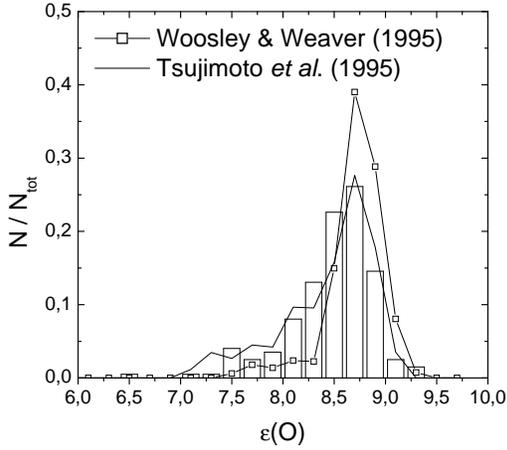}
  \caption{Distribution of oxygen abundances compared to model results, using two different
stellar evolution models.}
\end{figure}

\begin{figure}
  \includegraphics[height=.27\textheight]{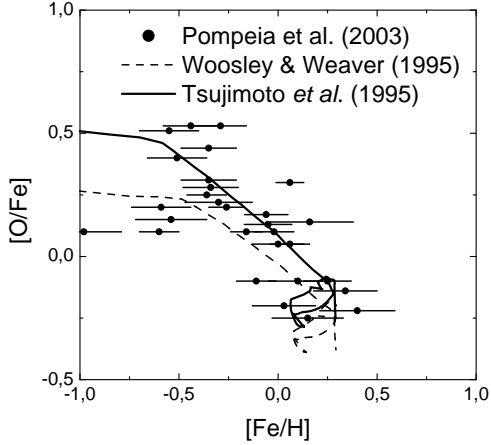}
  \caption{Evolution of the O/Fe ratio using two yields for SNII.}
\end{figure}

The first phase of the model consists in a fast collapse of primordial gas.
To allow a good reproduction of the derived chemical abundances found in low mass objects,
the adopted wind rate from supernovae is 50\% (40\% to the halo and 10\% to the
inner disk). Based on hydrodynamical simulations (Samland et al. 1997, Mac Low \& Ferrara 1998),
up to 50\% of the elements produced by supernovae can be ejected to the halo or inner disk.

\begin{figure}
  \includegraphics[height=.27\textheight]{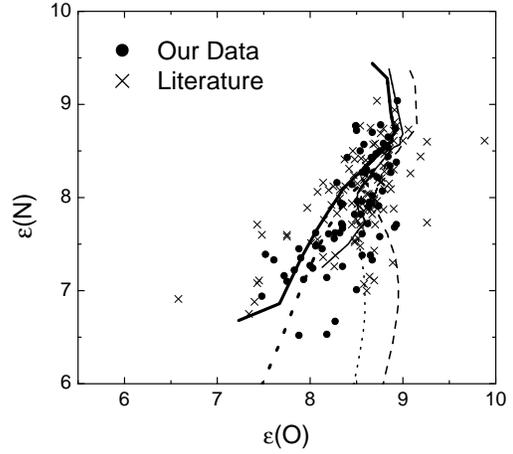}
  \caption{N/O ratio for our sample and data from the literature, compared to different model predictions (see text).}
\end{figure}

It can be seen from fig. 4 that there is a strong scattering in the derived oxygen abundances
for low nitrogen objects. This scattering can be reproduced assuming that the winds produced
by supernovae (II and Ia) eject part of the synthesized material. With this process, there
is simultaneously a chemical enrichment of the halo and inner disk and a decreasing in the
bulge enrichment; as a consequence, in the external regions the abundances of heavier 
elements such as oxygen and iron are 
increased with respect to nitrogen and helium, produced in intermediate mass stars.
 Bulge objects with high oxygen abundances and low nitrogen can also be found. 
 Their presence in the bulge indicates that they
were formed from a medium already oxygen-rich. Since oxygen originates from SNII, this population
was formed in an epoch when the interstellar medium was already enriched by ejecta
from short time-life stars, but still not enriched by elements formed in longer life stars
like nitrogen.

To have a low N/O ratio, the oxygen-rich gas would have to fall first in the inner disk,
before the beginning of its star-formation history, or in the halo region that later would
create the inner disk. If all the gas had fallen directly onto the bulge, its N/O ratio
could not explain values smaller than $log(N/O)=-1$. In a paper on the time variation of
the radial metalicity gradient of the Galaxy, Maciel et al. (2003) show that old objects
in the inner disk have high oxygen and low nitrogen abundances, indicating the occurrence
of a previous chemical enrichment for heavier elements.

Different sets of models are presented in fig. 4. Thick lines represent objects in the central
region and thin lines represent those in the external bulge-inner disk region. Therefore, 
the thin solid line represent objects
formed from gas already enriched by supernovae, and the thick solid line represents objects
formed by primordial gas, in the central region. 
In the same figure another model run is shown for both regions, adopting no oxygen and nitrogen yields for stars between
0.8 and 0.9 solar masses (dotted thick and thin lines). It can be seen that there is a considerable difference
between the curves for low abundance objects. This is due to the yields for low mass objects,
that can vary considerably. The dashed line was produced assuming that 25\% of material produced
by supernovae was ejected from the bulge to the inner disk. It should be kept in mind that
part of the scattering found in this figure can be due to observational uncertainties (Escudero
2001), as well as to the rotation of the progenitor star (Meynet \& Maeder 2002).

\section{Discussion}

The derived chemical abundance distribution for the observed objects is compatible with
other data from the literature. Mean values derived for the chemical species studied
are He/H=0.112, $\epsilon$(N)=7.83, $\epsilon$(O)=8.43, $\epsilon$(S)=6.56, $\epsilon$(Ar)=6.24,
$\epsilon$(Ne)=7.69, adopting $\epsilon$(X)=log(X/H)+12.

Results indicate that the best fit for the derived distribution of chemical abundances is achieved 
using a double-infall model, where the first one is a fast collapse of primordial gas 
and the second is slower and enriched by material ejected by the bulge itself during 
the first episode.

The galactic bulge had its first stellar formation episode triggered by
a large collapse of primordial gas. This episode is a consensus among all
galactic formation models, and is required based not only on chemical evolution
models but also on observational constraints. First, the large quantity of old
objects requires an important star formation episode at the beginning of
the evolution. Second, the wide distribution of metallicities observed in
stars and planetary nebulae can be better explained by a fast gas collapse
than by a slow one. Third, the presence of old metal-rich stars can be explained
if an abrupt gas collapse enriches rapidly the interstellar medium, generating
poor and rich old stars as well.

This first initial collapse has as main characteristic the large mass loss 
ejected by supernovae for external regions such as halo, disk or event to outside the Galaxy.
This mass loss is essential to reproduce the abundance distribution observed
in planetary nebulae. This is also a consensus among recent models for the 
galactic bulge: without this loss the stellar abundances would be higher than
those observed. For now, it is not possible to define exactly the amount of gas
lost in this episode. In this work, mass losses between 40 \% and 60 \% are suggested,
which are similar to those derived by Ferraras et al. (2003). A better definition
of this parameter requires more precise observational data, as well as more realistic
hydrodynamical models for the central region of the Galaxy.

The destination of the ejected material is not clearly understood yet. Samland et
al. (1997) propose that this material was ejected to the halo and later fell
on the disk. In this work they reproduce many chemical properties of the disk,
however one of their conclusions is that the disk abundance gradient begun to
form 6 Gyr after the beginning of the disk formation, which is contradictory with the time
variation of the abundance gradient derived by Maciel et al. (2003), which indicates
an already established gradient at this epoch.

Observational data from stars suggest that the IMF at the beginning of
bulge formation is more similar to Salpeter's than to Kroupa's, which means
that a large amount of massive objects was formed at the early phases
of the formation process, in a time scale of a few million years. In a later
phase, between 1 and 3 Gyr after the first collapse, the second gas collapse
occurred and formed the inner disk. 
This material was previously enriched by the gas ejected from the bulge
during the first collapse, or even by the chemical evolution of the halo.
This previous enrichment of the gas can explain the presence in the inner
disk of objects with high oxygen and low nitrogen abundances.

\begin{theacknowledgments}
  This work was partly supported by the Brazilian agencies FAPESP and CNPq.
\end{theacknowledgments}

\bibliographystyle{aipprocl} 

\bibliography{}


\end{document}